\documentclass[twocolumn,showpacs,groupedaddress]{revtex4}
\usepackage{graphicx}

\begin{document}

\newcommand{\ket}[1]{|#1\rangle}
\newcommand{\bra}[1]{\langle#1|}
\newcommand{\langlep}[1]{_p\langle}

\title{
Effect of feedback on the control of a two-level dissipative quantum
system}
\author{ L. C. Wang, X. L. Huang, X. X. Yi}

\affiliation{School of Physics and Optoelectronic Technology, Dalian
University of Technology, Dalian 116024, China}

\date{\today}

\begin{abstract}
We show that it is possible to modify the stationary state by a
feedback control in a two-level dissipative quantum  system. Based
on the geometric control theory, we also analyze the effect of the
feedback on the time-optimal control in the dissipative system
governed by the Lindblad master equation. These effects are
reflected in the function $\Delta_A(\vec{x})$ and
$\Delta_B(\vec{x})$ that characterize the optimal trajectories, as
well as the switching function $\Phi(t)$ and $\theta(t),$ which
characterize the switching point in  time for the time-optimal
trajectory.
\end{abstract}

\def\aL{ \hat{a}^{\dag}_{L} }
\def\aR{ \hat{a}^{\dag}_{R} }
\def\bL{ \hat{b}^{\dag}_{L} }
\def\bR{ \hat{b}^{\dag}_{R} }

\pacs{ 03.65.-w, 03.67.Mn,02.30.Yy } \maketitle

\section{Introduction}
The problem of controlling quantum systems has been an important
scientific and technological challenge \cite{blaquiere87} since the
discovery of quantum mechanics more than a century ago. Numerous
approaches to the control of a quantum system have been proposed in
the past decades, which depending on how the controls enter the
system, can be divided into two categories: the open loop scheme
(coherent control) and the closed loop scheme
\cite{huang83,peirce98,brockett83,romano07}. In the open loop scheme
the control functions are fixed, namely they can not be modified
according to the state of the quantum system. Whereas in the closed
loop scheme, the control functions are updated in real time by
feeding back some information about the actual state of the
system\cite{geremia04,reiner04,morrow02,bushev06}, this scheme is
called {\it quantum feedback control}
\cite{wiseman93,mancini98,belavkih99,doherty99}. Quantum feedback
control may emerge as a natural possible route to develop strategies
to prepare entangled states and prevent their
deterioration\cite{carvalho07} in controlled open
systems\cite{pechen06}.

For a real-world quantum system, the coupling of the system to its
environment is unavoidable. Because of this unavoidable coupling,
the system dynamics is subject to irreversibility, dissipation, and
dephasing. As a consequence, some appealing properties of quantum
systems, for example entanglement, are usually lost during the time
evolution, leading to many typical manifestations  of this
irreversibility, for example the relaxation of the system to a
stationary state. This fact naturally gives rise to the following
question, to what extent it is possible to modify the stationary
state by the controls? For open loop control, a previous study
\cite{romano07} shows that the stationary state can be modified by
indirect control through coupling the open system to an auxiliary
two-level system. What about the closed loop control? can a feedback
affects the stationary state of the system?

On the other hand, active research has been performed for
manipulating an open system with more realistic situations taken
into account. Although the controls can not fully compensate the
effect of decoherence for an open system governed by the Lindblad
master equation, as shown in Ref. \cite{altafini04}, an efficient
control can still be achieved\cite{jirari05,wenin06,sugny06}. These
analyses are based on the numerical optimalization techniques, and
it seems that only controls can be achieved by these numerical
methods. By the geometric control theory, quantum system with few
levels (e.g., two- or three-level systems) can be formulated
\cite{stefanatos04,sugny07} analytically. This analysis
\cite{sugny07} is for the time-optimal control of a dissipative
two-level quantum system without  feedback. In this paper we put
forward the study of the control by taking the feedback into
account, a two-level system governed by the Lindblad master equation
will be chosen to detail the analysis.

The paper is organized as follows. In Sec.{\rm II}, we specify the
dynamical settings considered in this paper, and study the effect of
feedback control on the stationary state of the system. By using the
Pontryagin maximum principle, the time-optimal control of a
two-level dissipative system with a feedback control will be studied
in Sec.{\rm III}.  Conclusion and discussion are presented in
Sec.{\rm IV}. Analytical solutions to the system are given in the
Appendix.

\section{Relaxation to stationary states of
a two-level dissipative system with feedback}

In this section, we shall show that a Markovian feedback scheme
based on the continuous monitoring of quantum jumps, can lead to an
improvement of control of the stationary states. Before
investigating the influence of feedback on the stationary states,
let us first briefly analyze the case without feedback. To start
with, we consider a two-level system with the free Hamiltonian $H_0$
and the control Hamiltonian $H_1$ (with a control field $u$). Within
the Markov approximation for the system-environment interaction, the
time evolution of the two-level system is described by the Lindblad
master equation, $ \frac{\partial}{\partial t}\rho=-i[H, \rho]
+\Gamma(\sigma^-\rho\sigma^+-\frac 1 2\sigma^+\sigma^-\rho-\frac 1
2\rho\sigma^+\sigma^-) +\gamma(\sigma^+\rho\sigma^- -\frac 1
2\sigma^-\sigma^+\rho-\frac 1 2\rho\sigma^-\sigma^+). $ Here $
H=H_0+uH_1$\cite{noteh0}; $\sigma_z=|e\rangle\langle
e|-|g\rangle\langle g|,$ $\sigma^-=|g\rangle\langle e|$ and
$\sigma^+=|e\rangle\langle g|$ are the Pauli matrices.
$\Gamma=(\bar{n}+1)\kappa$ and $\gamma=\bar{n}\kappa.$ We denote by
$\bar{n}$ the noise intensity of the environment, and $\kappa$ the
spontaneous emission rate of the two-level system. $|e\rangle$ and
$|g\rangle$ stand for the excited and ground states of the two-level
system, respectively. Choosing $uH_1=u^*|g\rangle\langle
e|+u|e\rangle\langle g|$, we obtain the stationary state of the
dissipative two-level system (setting $u=u_1+iu_2$)
$\rho_{\infty}=\frac 1 2 \left (\matrix{ 1-x_3 & x_1+ix_2\cr
x_1-ix_2 & 1+x_3 }\right)$, where
\begin{eqnarray}
x_1&=&\frac{4u_2x_3}{\Gamma+\gamma} ,\nonumber\\
x_2&=& -\frac{4u_1x_3}{\Gamma+\gamma},\nonumber\\
x_3&=&\frac{(\Gamma^2-\gamma^2 )} {8|u|^2 +
(\gamma+\Gamma)^2},\label{ss1}
\end{eqnarray}
and $\vec{x}=(x_1,x_2,x_3)$ was defined by
\begin{eqnarray}
x_1&=&2\texttt{Re}\rho_{eg}, \nonumber\\
x_2&=&2\texttt{Im}\rho_{eg}, \nonumber\\
x_3&=&\rho_{gg}-\rho_{ee}.\label{notationx}
\end{eqnarray}
\begin{figure}
\includegraphics*[width=0.7\columnwidth,
height=0.5\columnwidth]{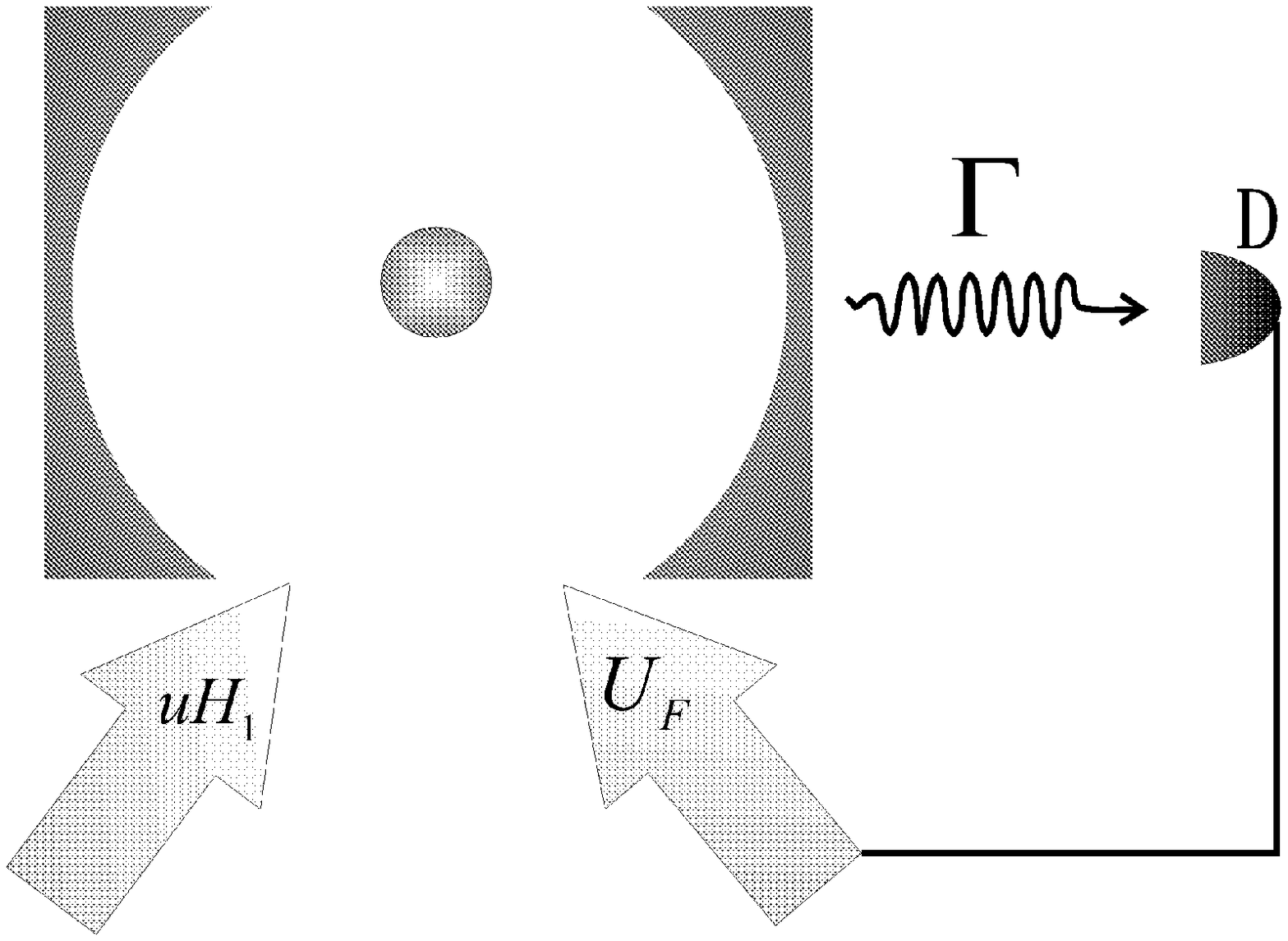} \caption{(Color online)Schematic
illustration of the feedback control and the open loop control. The
system consists of an atom simultaneously driven by $uH_1$ and a
feedback $U_F$. This feedback is conditioned on the measurement of
the output of the leaky cavity.} \label{fig1}
\end{figure}
Two observations can be made from Eq.(\ref{ss1}). (1) The control
field $u$ drastically changes the stationary state, and $x_1=x_2=0$
when $u=0;$ (2) $x_2$ and $x_1$ are proportional to $x_3$,
considering $x_1^2+x_2^2+x_3^2\leq 1,$ we thus have $x_3^2\leq
\frac{(\Gamma+\gamma)^2}{(\Gamma+\gamma)^2+16|u|^2}.$ Therefore it
is possible to manipulate the stationary state by using the open
loop control $uH_1,$ but it can not compensate fully the decoherence
effect, for instance with specific $\Gamma$ and $\gamma$, we can not
reach all states through relaxing the system to its equilibrium.

Taking the cavity QED system as an example,  we now introduce the
description of the measurement and feedback scheme, this  is
schematically shown in figure \ref{fig1}. The cavity output is
monitored by a photon detector $D$ whose signal provides the input
to the application of the closed-loop control $U_F$. Note that we
have two controls in this scheme, one is the open-loop scheme
corresponding to the control Hamiltonian $uH_1$, and another is the
closed-loop control denoted by $U_F$. The feedback control $U_F$ is
triggered immediately only after a detection click, namely a quantum
jump occurs. This scheme was used to generate and protect entangled
steady state in cavity QED system \cite{carvalho07}. The master
equation for our system reads \cite{wiseman93},
\begin{eqnarray}
\frac{\partial}{\partial t}\rho&=&-i[H, \rho]
+\Gamma(U_F\sigma^-\rho\sigma^+U^{\dag}_F-\frac 1
2\sigma^+\sigma^-\rho-\frac 1 2\rho\sigma^+\sigma^-)\nonumber\\
&+&\gamma(\sigma^+\rho\sigma^--\frac 1 2\sigma^-\sigma^+\rho-\frac 1
2\rho\sigma^-\sigma^+).\label{me}
\end{eqnarray}
The jump feedback $U_F\sigma^-\rho\sigma^+U_F^{\dag}$ can be
understood as follows. The unitary operator $U_F$ is applied only
immediately after a detection event, which is described by the term
$\sigma^-\rho\sigma^+.$ Intuitively the stationary states depend on
the feedback operator $U_F.$  So, once the measurement prescription
has been chosen, the freedom to design a feedback to produce a
stationary state lies in the different choices for the feedback
operator $U_F.$ Although an enormous range of possibilities for
$U_F$ is allowed, even considering the limitations imposed by
experimental constraints,  we here choose (with the constraint
$U_FU_F^{\dagger}=1$)
\begin{eqnarray}
U_F=a\sigma_x+b\sigma_y+c\sigma_z.
\end{eqnarray}
In fact the feedback $U_F$ written in this form covers all allowed
possibilities. By Setting $\dot{\rho}=0$, we arrive at the
stationary state,
\begin{widetext}
\begin{eqnarray}
x_3&=&\frac{4\Gamma(-u_1f_2+u_2f_1)+(\gamma+\Gamma)(\gamma-\Gamma
|U_F^{gg}|^2)}
{4u_1(-2u_1-\Gamma f_2)-4u_2(2u_2-\Gamma f_1)-(\gamma+\Gamma)(\gamma+\Gamma |U_F^{gg}|^2)},\nonumber\\
x_1&=&\frac{(4u_2-2\Gamma f_1)x_3+2\Gamma f_1}{\Gamma+\gamma} ,\nonumber\\
x_2&=& \frac{(-4u_1-2\Gamma f_2)x_3+2\Gamma
f_2}{\Gamma+\gamma},\label{ss2}
\end{eqnarray}
\end{widetext}
where $f=U_F^{eg}U_F^{gg}=f_1+if_2$ was defined with
$U_F^{ij}=\langle i|U_F|j\rangle, i,j=e,g.$ This stationary state
differs from that in the case  without feedback at (1)$x_1$ and
$x_2$ are not zero even if $u=0$; (2) $x_1$ and $x_2$ are not
proportional to $x_3$; (3) $x_1^2+x_2^2+x_3^2$, i.e., the trace of
$\rho_{\infty}^2$ ($\rho_{\infty}$ is the stationary state) does not
only depend on $|u|^2$. This indicates that we may change the
reachable set of stationary states by the feedback control.
Mathematically, we have only two independent parameters in
Eq.(\ref{ss1}), when $\Gamma$ and $\gamma$ are fixed, whereas there
are (at least) three degrees of freedom in Eq.(\ref{ss2}), matching
(beyond) the number of independent parameter in the two-level
system.

\section{Time optimal control of the system with Markovian feedback}
\begin{figure}
\includegraphics*[width=0.8\columnwidth,
height=0.6\columnwidth]{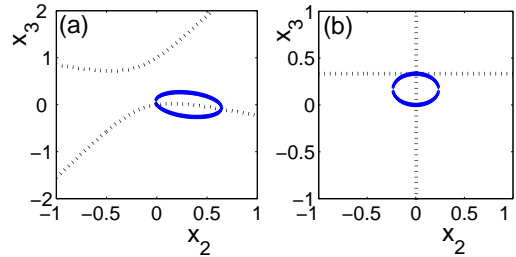} \caption{The sub-manifold of
$(x_2,x_3)$ is divided by $C_A=\Delta_A^{-1}(0)$(solid lines) and
$C_B=\Delta^{-1}(0)$ (dashed lines). $\Gamma=0.6, \gamma=0.3$ were
chosen for this plot. (a) and (b) are for the case with and without
feedback, respectively.} \label{fig2}
\end{figure}

\begin{figure}
\includegraphics*[width=0.8\columnwidth,
height=0.6\columnwidth]{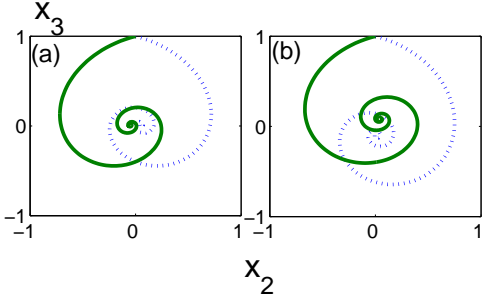} \caption{Analytical solutions to
the dynamics with $u=-1$(solid line) and $u=1$ (dashed line). (a)
Without feedback, and (b) with feedback. The other parameters chosen
are $\Gamma=0.4$,$\gamma=0.3$, and $\beta=0.2\pi.$} \label{fig3}
\end{figure}

Optimal control theory has a long-standing tradition in various
fields of physics \cite{bryson75}. To our knowledge, one of the
first application to a quantum system has been in the field of
quantum chemistry \cite{peirce98}. Recently, optimal control theory
has been extended to dissipative systems
\cite{sugny07,hohenester04,jirari05}. In this section, we consider
the time-optimal control of a dissipative two-level system described
by the master equation (\ref{me}). This problem was studied in
Ref.\cite{sugny07} without feedback, here we will focus on the
effect of the feedback on the time optimal control of this system.
With the  notations in Eq.(\ref{notationx}), the master equation can
be written as,
\begin{eqnarray}
\dot{x_1}&=&2u_2x_3-\frac{\Gamma+\gamma}2x_1+\Gamma f_1(1-x_3),\nonumber\\
\dot{x_2}&=&-2u_1x_3-\frac{\Gamma+\gamma}2x_2+\Gamma f_2(1-x_3),\nonumber\\
\dot{x_3}&=&2u_1x_2-2u_2x_1-x_3(\gamma+\Gamma|U_F^{gg}|^2)-
(\gamma-\Gamma|U_F^{gg}|^2).\nonumber\\
\end{eqnarray}
As defined in Eq.(\ref{notationx}), $x_1,x_2,x_3$ are three real
parameters, while $u$ is a complex function $u=u(t)=u_1(t)+iu_2(t),$
and $f=U_F^{eg}U_F^{gg}=f_1+if_2$ is a complex number. As the most
important tool for the study of optimal control, the Pontryagin
maximum principle (PMP) provides a first order necessary condition
for optimality. In the following we shall analyze the optimal
control of this two-level system by applying the PMP. The analysis
of the optimal control on R$^3$  manifold is considerably
complicated.  To simplify the study, we restrict the dynamics to a
sub-manifold R$^2$  by assuming the control field $u$ is real and
$f_1=0$ \cite{note1}. With these assumptions, $x_2$ and $x_3$ are
decoupled from $x_1$, leading to
\begin{eqnarray}
\dot{x_2}&=&-2u_1x_3-\frac{\Gamma+\gamma}2x_2+\Gamma f_2(1-x_3),\nonumber\\
\dot{x_3}&=&2u_1x_2-(\gamma+\Gamma|U_F^{gg}|^2)x_3-(\gamma-\Gamma|U_F^{gg}|^2).
\label{x2}
\end{eqnarray}
To shorten the notation, we omit the index $1$ of $u_1$ and set
$\vec{x}=(x_2,x_3).$ Equation (\ref{x2}) becomes
 $\dot{\vec{x}}=F+uG$,
 where
\begin{eqnarray}
F=\left(
    \begin{array}{c}
      -\frac{\Gamma+\gamma}2x_2+\Gamma f_2-\Gamma f_2x_3 \\
      -(\gamma+\Gamma|U_F^{gg}|^2)x_3-(\gamma-\Gamma|U_F^{gg}|^2) \\
    \end{array}
  \right)
\end{eqnarray}
and
\begin{eqnarray}
G=\left(
    \begin{array}{c}
      -2x_3 \\
      2x_2 \\
    \end{array}
  \right)
\end{eqnarray}
For each $\vec{x}$, we have calculated $\Delta_A(\vec{x})=Det(F,G)$
and $\Delta_B(\vec{x})=Det(G,[F,G])$ defined in \cite{boscain04},
\begin{eqnarray}
\Delta_A(\vec{x})&=&-2(\gamma+\Gamma|U_F^{gg}|^2)x_3^2-
2(\gamma-\Gamma|U_F^{gg}|^2)x_3\nonumber\\
&-&(\gamma+\Gamma)x_2^2-2\Gamma f_2x_2x_3+2\Gamma f_2x_2,\nonumber\\
\Delta_B(\vec{x})&=&4(\gamma-\Gamma|U_F^{gg}|^2)x_2
+4(\gamma-\Gamma+2\Gamma|U_F^{gg}|^2)x_2x_3\nonumber\\
&+&4\Gamma f_2x_2^2-4\Gamma f_2x_3^2+4\Gamma f_2x_3,\label{delta}
\end{eqnarray}
where $\Delta_A(\vec{x})$ is useful for studying abnormal extremals,
and $\Delta_B(\vec{x})$ is for detecting singular trajectories. We
can find from Eq.(\ref{delta}) that the feedback control $U_F$ play
an important role in $\Delta_A(\vec{x})$ and $\Delta_B(\vec{x})$,
which are crucial in the time-optimal problem. We denote by $C_A$
and $C_B$ the two sets of points $\Delta_A^{-1}(0)$ and
$\Delta_A^{-1}(0)$, respectively. $C_A$ and $C_B$ are responsible
for quantitative modification of the optimal trajectories. For
$U_F=1$ (without feedback), $\Delta_A(\vec{x})$ and
$\Delta_B(\vec{x})$ reduce to
\begin{eqnarray}
\Delta_A(\vec{x})&=&-2(\gamma+\Gamma)x_3^2- 2(\gamma-\Gamma)x_3
-(\gamma+\Gamma)x_2^2,\nonumber\\
\Delta_B(\vec{x})&=&4(\gamma-\Gamma)x_2
+4(\gamma+\Gamma)x_2x_3.\label{delta1}
\end{eqnarray}
In this case the set $C_B$ consists of the following two lines,
$x_2=0$ and $x_3=(\Gamma-\gamma)/(\gamma+\Gamma),$ while the
solutions of the  polynomial equation $2(\gamma+\Gamma)x_3^2+
2(\gamma-\Gamma)x_3 +(\gamma+\Gamma)x_2^2=0$ belong to the set $C_A$
(as shown in figure \ref{fig2}-(b)).  The two sets $C_A$ and $C_B$
change when the feedback is added to the system. This is illustrated
in figure \ref{fig2}-(a), where
we plot $C_A$ and $C_B$ with $\Gamma=0.6$, $\gamma=0.3$, $f=\cos
\frac\pi 5 \sin \frac\pi 5$, and $U_{F}^{gg}=\cos^2 \frac \pi 5$
\cite{note1}. Without feedback, the sub-manifold of R$^2$ is
symmetrically divided by the line $x_2=0$, as figure \ref{fig1}-(b)
shows. This symmetry is broken by the feedback (see figure
\ref{fig1}-(a)), indicating that the symmetry in the optimal
trajectory disappears. This is confirmed by figure \ref{fig3} where
the two trajectories with $u=\pm1$ are plotted.
\begin{figure}
\includegraphics*[width=0.8\columnwidth,
height=0.6\columnwidth]{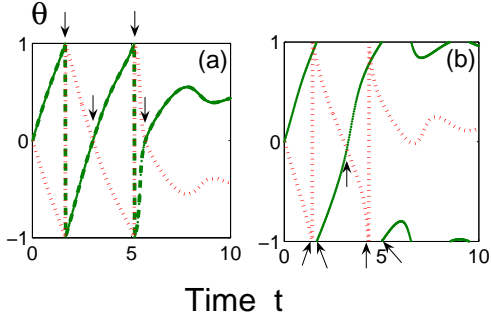} \caption{Illustration of
$\theta(t)$ as a function of time. The solid line and dashed line
represent the case of $u=-1$ and $u=1$, respectively. (a) Without
feedback, while (b) with feedback. The other parameters are the same
as in figure \ref{fig3}.} \label{fig4}
\end{figure}
Assuming the field $u$ is bounded by $|u| \leq 1,$ we now analyze
the optimal control of the two-level dissipative system with the
constraint of minimizing the total time of the control, i.e.,
time-optimal control. The Pontryagin maximum principle tells us that
for the model system considered here, the extremal field $u$ may
take either $-1$ or $1$ according to $u=sgn[\Phi(t)]$, if
$\Phi(t)\neq 0,$ where
\begin{eqnarray}
\Phi(t)=\vec{p}\cdot G=-2p_2x_3+2p_3x_2
\end{eqnarray}
with $\vec{p}$ satisfying
\begin{eqnarray}
\dot{p_2}&=&-\frac{\partial H}{\partial x_2}=\frac{\Gamma+\gamma}2p_2-2up_3,\nonumber\\
\dot{p_3}&=&-\frac{\partial H}{\partial x_3}=2up_2+\Gamma
f_2p_2+(\gamma+\Gamma|U_F^{gg}|^2)p_3.
\end{eqnarray}
If $\Phi(t)$ vanishes on an interval $[t_0,t_1]$, the corresponding
control in this interval is $u=\phi$, where $\phi$ can be calculated
by
\begin{eqnarray}
\frac d{dt}\Delta_B=\frac{\partial \Delta_B}{\partial x_2}\dot{x_2}
+\frac{\partial \Delta_B}{\partial x_3}\dot{x_3}=0,
\end{eqnarray}
leading to
\begin{eqnarray}
u=\frac{K_{33}x_3+K_{23}x_2x_3+K_{22}x_2^2+K_3x_3+K_2x_2}
{K'_{33}x_3+K'_{23}x_2x_3+K'_{22}x_2^2+K'_3x_3+K'_2x_2},
\end{eqnarray}
in our case. Here
\begin{eqnarray}
K_{33}&=&-(\Gamma+\gamma)\Gamma f_2, \nonumber\\
K_{23}&=&(\gamma-\Gamma+2\Gamma |U_F^{gg}|^2)(\frac{\Gamma}2+\frac{3\gamma}2+\Gamma |U_F^{gg}|^2)+2\Gamma^2f_2^2, \nonumber\\
K_{22}&=&(\Gamma+\gamma)\Gamma f_2, \nonumber\\
K_3&=&(\Gamma-\gamma)\Gamma f_2,\nonumber\\
K_2&=&(\gamma-\Gamma|U_F^{gg}|^2)(-\frac{\Gamma}2+\frac{3\gamma}2+2\Gamma|U_F^{gg}|^2)-2\Gamma^2f_2^2,\nonumber\\
K'_{33}&=&-2(\gamma-\Gamma+2\Gamma|U_F^{gg}|^2), \nonumber\\
K'_{23}&=&-8\Gamma f_2, \nonumber\\
K'_{22}&=&2(\gamma-\Gamma+2\Gamma|U_F^{gg}|^2) , \nonumber\\
K'_3&=&-2(\gamma-\Gamma|U_F^{gg}|^2),\nonumber\\
K'_2&=&2\gamma f_2.\nonumber\\
\end{eqnarray}
Notice that if $\Phi(t)$ has no zeros then $u$ is almost everywhere
constantly equal to $\pm 1$. Hence we are interested in determining
when the control may change sign. This problem can be studied in two
different ways: either by means of the switching function $\Phi(t)$
(as above) or using the function $\theta(t).$ Here $\theta(t)$ is
defined as the angle of rotation of the adjoint vector
$\vec{v}=(v_2,v_3)=(\dot{x}_2,\dot{x}_3)$ with respect to its
initial position. By the definition of $\vec{v}$, we obtain
\begin{eqnarray}
\dot{v}_2&=&-(2u+\Gamma
f_2)v_3-\frac{\Gamma+\gamma}{2}v_2,\nonumber\\
\dot{v}_3&=&2uv_2-(\gamma+\Gamma|U_F^{gg}|^2)v_3,
\end{eqnarray}
with the initial conditions
$v_2(0)=-2ux_3(0)-0.5(\Gamma+\gamma)x_2(0)+\Gamma f_2(1-x_3(0))$ and
$v_3(0)=2ux_2(0)-(\gamma+\Gamma|U_F^{gg}|^2)x_3(0)-(\gamma-\Gamma|U_F^{gg}|^2).$

In figure \ref{fig4}, we illustrate the angle $\theta(t)$ as a
function of time for the system without (figure (a)) and with
(figure (b)) feedback. The times at which the control $u$ can switch
are marked by arrows.
\begin{figure}
\includegraphics*[width=0.8\columnwidth,
height=0.6\columnwidth]{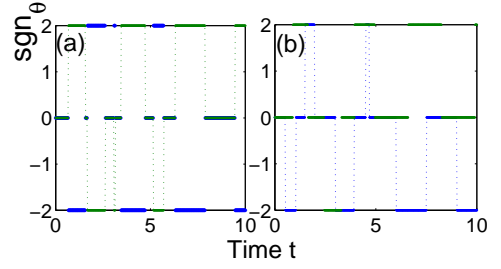}\caption{$sgn_{\theta}$ versus
time $t$. The parameters chosen are the same as in figure
\ref{fig4}. Times corresponding to $sgn_{\theta}=0$ are allowed to
switch. This together with figure \ref{fig4} determine the switching
points in time.} \label{fig5}
\end{figure}
We found from figure \ref{fig4}-(a) that the two trajectories
corresponding to $u=\pm 1$ may switch at same times when there is no
feedback, resulting in symmetry in the optimal trajectories. This
result is changed by the feedback as figure \ref{fig4}-(b) shows,
the two trajectories switch almost at different times. In addition,
the trajectories can switch to apposite control when $\theta > 0$
and $\dot{\theta} > 0$ or $\theta< 0$ and $\dot{\theta}< 0$, this
condition together with the results in figure \ref{fig4} yield the
switching points in time. Define
$sgn_{\theta}=sgn(\theta)-sgn(\dot{\theta})$, the allowed switching
points in terms of $sgn_{\theta}$ are plotted in figure \ref{fig5}.
\section{conclusion}
For a dissipative two-level system, we have shown that the
manipulation of  stationary state by feedback is possible. The
dependence of the stationary state on the feedback has been
calculated and discussed. The feedback together with the open loop
control can broaden the reachable set of the stationary state, which
provides us a new method to prepare quantum states by relaxing the
system to its equilibrium states. In addition to improve the control
of stationary state, the feedback also affects the time-optimal
control of a two-level dissipative quantum system. The optimal
trajectories for the system with feedback are no longer symmetric,
indicating that the optimal trajectory can not be determined by
using symmetry analysis. The switching points in time are also
presented and discussed, which together with the analytical
solutions to the dynamics in the Appendix yield the optimal
trajectory for the open system.

\section*{ACKNOWLEDGEMENTS} This work was supported by  NSF of China
under grant No. 60578014 and No. 10775023.

\appendix
\section*{APPENDIX:  analytical determination of the dynamics}

In this section, we solve analytically the dynamics of the system
given by Eq.(\ref{x2}). To simplify the expression, we rewrite the
equations as
\begin{eqnarray}
\dot{x_2}=A_2x_2+B_2x_3+C_2,\nonumber\\
\dot{x_3}=A_3x_2+B_3x_3+C_3,\label{d1}
\end{eqnarray}
where $A_2=-\frac{\Gamma+\gamma}{2}, B_2=-2u-\Gamma f_2,$
$C_2=\Gamma f_2,$ $A_3=2u, B_3=-(\gamma+\Gamma|U_F^{gg}|^2),$ and
$C_3=-(\gamma-\Gamma|U_F^{gg}|^2).$ We assume that $u=\pm 1$, simple
algebra gives the exact solutions to Eq. (\ref{d1}),
\begin{eqnarray}
x_2(t)=L'+M'e^{\alpha_2t}+N'e^{\alpha_3t},\nonumber\\
x_3(t)=L+Me^{\alpha_2t}+Ne^{\alpha_3t},
\end{eqnarray}
where
\begin{widetext}
\begin{eqnarray}
\alpha_2,\alpha_3=\frac{A_2+B_3}2\pm\frac 1
2\sqrt{(A_2+B_3)^2-4(A_2B_3-A_3B_2)},
\end{eqnarray}
\begin{eqnarray}
L'&=&\frac{C_3B_2-C_2B_3}{A_2B_3-A_3B_2},\nonumber\\
M'&=&\frac{L'(A_2+B_3)+B_2x_3(0)-B_3x_2(0)+C_2+(x_2(0)-L')
\alpha_2}{\alpha_2-\alpha_3},\nonumber\\
N'&=&x_2(0)-L'-M',
\end{eqnarray}
\begin{eqnarray}
L&=&\frac{A_3C_2-A_2C_3}{A_2B_3-A_3B_2},\nonumber\\
M&=&\frac{L(A_2+B_3)-A_2x_3(0)+A_3x_2(0)+C_3+(x_3(0)-L)\alpha_2}
{\alpha_2-\alpha_3},\nonumber\\
N&=&x_3(0)-L-M,
\end{eqnarray}
\end{widetext}


\begin{references}
\bibitem{noteh0} By choosing two degenerate levels and setting the
energy  to be zero, we may safely drop $H_0$ in the discussion.
Alternatively, the situation discussed here can be understood as a
case where the cavity field resonantly couples to the two-level
system, and let $u=u(t)exp^{i\omega t}$ with $\omega$ being the
level difference(also the frequency of the cavity field). In the
interaction picture, we can have the   results in the context.
\bibitem{note1} This is the case for $a=0$, $b=\sin\beta$
and $c=\cos\beta$ in $U_F,$ where $\beta\in [0,2\pi).$ Clearly
$U_F^{\dagger}U_F=1$ satisfies the constraint. $x_1(0)$ may be set
to be zero, hence $x_1(t)=0.$
\end{references}
\end{document}